\documentclass[prb, twocolumn]{revtex4-2}

\usepackage{graphicx}
\usepackage{dcolumn}
\usepackage{bm}
\usepackage{hyperref}
\usepackage{xcolor}
\usepackage{mathrsfs}
\usepackage{enumitem} 
\usepackage{placeins}
\usepackage{academicons}
\usepackage{orcidlink} 
\usepackage{amsmath}
\usepackage{amsfonts}
\usepackage[normalem]{ulem}
\usepackage{physics}
\usepackage{tcolorbox}
\usepackage{subcaption}
\usepackage{ragged2e}

\definecolor{notecolor}{RGB}{220, 20, 60} 
\definecolor{editcolor}{RGB}{30, 144, 255} 
\definecolor{improvementcolor}{RGB}{34, 139, 34} 

\begin{document}

\preprint{APS/123-QED}

\title{
Stabilizer entropy in non-integrable quantum evolutions
}

\author{J. Odavi\'{c}\orcidlink{0000-0003-2729-8284}$^{1,2}$}
\email[Corresponding author: ]{jovan.odavic@unina.it}
\author{M. Viscardi\orcidlink{0000-0002-1606-9622}$^{1,2}$}
\author{A. Hamma\orcidlink{0000-0003-0662-719X}$^{1,2,3}$}

\affiliation{$^1$Dipartimento di Fisica Ettore Pancini, Universit\`a degli Studi di Napoli Federico II, via Cinthia, 80126 Fuorigrotta, Napoli, Italy}
\affiliation{$^2$INFN, Sezione di Napoli}
\affiliation{$^3$Scuola Superiore Meridionale, Largo S. Marcellino 10, 80138 Napoli, Italy}

\date{\today}

\begin{abstract}
Entanglement and stabilizer entropy are both involved in the onset of complex behavior in quantum many-body systems. Their interplay is at the root of complexity of simulability, scrambling, thermalization and typicality. In this work, we study the dynamics of entanglement, stabilizer entropy, and the anti-flatness of the entanglement spectrum after a quantum quench in a spin chain. We find that free-fermion theories show a gap in the long-time behavior of these resources compared to their random matrix theory value while non-integrable models saturate it.

\end{abstract}

\maketitle

\section{\label{sec:introduction} Introduction}
Understanding quantum chaos in many-body systems is crucial to a variety of scientific fields. In statistical physics, the onset of quantum chaos is closely tied to the mechanisms of thermalization~\cite{Rigol_Dunjko_Olshanii_2008, Srednicki_1994, DAlessio_Kafri_Polkovnikov_Rigol_2016}. In high-energy physics, chaos and information scrambling play a fundamental role in understanding black hole dynamics~\cite{Sekino_Susskind_2008, Maldacena_Shenker_Stanford_2016, Dowling_Kos_Modi_2023, PhysRevLett.126.030601, Leone_Oliviero_Zhou_Hamma_2021}. Interestingly, quantum chaos is also relevant to the concept of quantum advantage—the demonstrable speed-up of algorithms run on quantum hardware. In quantum information science, fully programmable quantum computers derive their power from their universality, or their ability to explore the full Hilbert space ergodically~\cite{DiVincenzo_2000, True_Hamma_2022}. The ability to harness ergodicity—and with it, quantum chaos—lies at the core of quantum computing’s transformative potential~\cite{Daley_Bloch_Kokail_Flannigan_Pearson_Troyer_Zoller_2022, Fauseweh_2024, Berke_Varvelis_Trebst_Altland_DiVincenzo_2022, Borner_Berke_DiVincenzo_Trebst_Altland_2024}. 

While entanglement is widely regarded as a key factor in the efficient classical simulation of quantum systems and their dynamics~\cite{Schollwock_2011, Orus_2014}, it alone does not guarantee a quantum speed-up over classical algorithms. In fact, there are highly entangled quantum states that remain classically tractable. A prominent example is the class of stabilizer states, generated by operations from the Clifford group—a finite subgroup of the unitary group~\cite{Gottesman_1998, Aaronson_Gottesman_2004, Veitch_Mousavian_Gottesman_Emerson_2014}. Despite their high entanglement, these states can still be efficiently simulated using classical methods. Indeed, the appearance of quantum chaos can be seen as the onset of quantum complexity in the entanglement pattern that results from injecting - through quantum circuit doping or measurements of non-Clifford resources~\cite{Oliviero_Leone_Hamma_2021, True_Hamma_2022, Zhou_Yang_Hamma_Chamon_2020, Hinsche_Ioannou_Nietner_Haferkamp_Quek_Hangleiter_Seifert_Eisert_Sweke_2023, Bejan_McLauchlan_Beri_2024}.

In quantum circuit dynamics, it was demonstrated that states generated by the Clifford group fail to exhibit true quantum chaotic behavior. This limitation arises because the Clifford group forms only a 3-design, rendering it non-universal~\cite{Gross_2016, Roberts_Yoshida_2017}. The concept of a $t$-design refers to how closely an ensemble of unitaries approximates the first $t$ moments of an ensemble sampled according to the Haar measure. To achieve higher degrees of design, gates outside the Clifford group - such as T-gates - are introduced, which generate non-stabilizer states and promote the circuit dynamics to a higher complexity~\cite{Zhou_Yang_Hamma_Chamon_2020, Leone_Oliviero_Zhou_Hamma_2021, Hinsche_Ioannou_Nietner_Haferkamp_Quek_Hangleiter_Seifert_Eisert_Sweke_2023}. More specifically, complexity is reflected in the fact that the circuit simulation times grow exponentially in the system size. Whereas in the case of states generated via the application of gates that belong to the Clifford group grow only polynomially and thus can be efficiently simulated via classical computers~\cite{ Aaronson_Gottesman_2004, Bravyi_Kitaev_2005}. Due to these properties, non-stabilizer states — those that transcend the Clifford framework and move towards the universality and chaos characteristic of generic states in the vast Hilbert space — are referred to as ``magic states''.

The insights from quantum circuits naturally lead to the following question: \textit{How does quantum chaos emerge in quantum dynamics governed by local Hamiltonian evolution?} 
This question arises as it is well established that quantum chaos in quantum dynamics generated by a local Hamiltonian arises from the non-integrability of the Hamiltonian~\cite{DAlessio_Kafri_Polkovnikov_Rigol_2016}. Does this imply that integrable systems exhibit dynamics analogous to those generated by Clifford circuits? The answer, as addressed in recent studies~\cite{Oliviero_Leone_Hamma_2022, Rattacaso_Leone_Oliviero_Hamma_2023}, is a resounding no. States of integrable Hamiltonian many-body systems possess extensive non-stabilizerness reflected in the linear scaling with the qubit/spin number~\cite{Liu_Winter_2022}.  Evolving such states via integrable quantum protocols leads to a further increase in such resources~\cite{Rattacaso_Leone_Oliviero_Hamma_2023}. Thus, there is no straightforward analogy between circuit dynamics and Hamiltonian evolution.

In this work, we provide a more comprehensive exposition and comparisons between integrable and non-integrable dynamics in a closed quantum system through the lens of non-Clifford resources and their interplay with entanglement production. Specifically, we examine the long-time behavior of quantum systems subjected to a sudden quench, driving them out of equilibrium. Such protocols are commonly encountered in cold atom experiments and solid-state platforms, where understanding their dynamics is crucial for demonstrating quantum advantage~\cite{Daley_Bloch_Kokail_Flannigan_Pearson_Troyer_Zoller_2022, Fauseweh_2024,Kim_Wood_Yoder_Merkel_Gambetta_Temme_Kandala_2023}

As previously mentioned, entanglement alone is insufficient to induce the full complexity observed in quantum many-body systems, nor does it guarantee the systems will exhibit universal behavior. Therefore, our focus extends beyond entanglement to the resources outside Clifford (or stabilizer) framework~\cite{Gottesman_1998}. In particular, we employ the Stabilizer Rényi Entropy (SRE)~\cite{Leone_Oliviero_Hamma_2022} to quantify the non-stabilizerness generated by these quantum quench protocols. SRE quantifies the extent to which a quantum state spreads over the Pauli operator basis and serves as a computationally tractable measure of non-stabilizerness, which is considered a crucial resource in quantum computation.

In addition to entanglement and SRE, we explore the anti-flatness of the entanglement spectrum~\cite{Tirrito_Tarabunga_Lami_Chanda_Leone_Oliviero_Dalmonte_Collura_Hamma_2024, Odavic_Torre_Mijic_Davidovic_Franchini_Giampaolo_2023}. Anti-flatness quantifies how much the entanglement spectrum, defined as the eigenvalues of the reduced density matrix across a bipartition, deviates from a flat or uniform distribution. Stabilizer states are characterized by a flat entanglement spectrum, so any deviation from flatness captures non-stabilizer features encoded in the spectrum and provides further insight into state complexity. Our results reveal a strong correlation between the behavior of the SRE for the total state and the anti-flatness of the entanglement spectrum within subsystems, offering new perspectives on the interplay between these two quantities.

The relationship between global and local properties (defined on a subsystem) of quantum systems is of central importance to understanding quantum thermalization~\cite{Rigol_Dunjko_Olshanii_2008, Srednicki_1994, DAlessio_Kafri_Polkovnikov_Rigol_2016}. In classical physics, thermal equilibration occurs when a physical system evolves into a steady state in which observable quantities no longer change significantly over time. This process reflects an inherent loss of memory about the system's initial conditions.

In contrast, closed quantum systems evolve unitarily, meaning that their time evolution is reversible and retains complete information about the initial state. In such systems, thermalization arises not from loss of information, but from the practical inaccessibility of that information through local measurements. That is, information initially localized in a small region of the system becomes spread out or  ``scrambled'' across many degrees of freedom. Once this scrambling occurs, local subsystems—described by reduced density matrices obtained by tracing out the rest of the system—appear thermal, and cannot distinguish between a truly thermal state and a highly entangled pure state. This raises questions about the conditions under which quantum states evolve into long-time behavior resembling random pure states (considered typical), and how the nature of the dynamics and initial conditions affect this evolution.

In the evolution of closed quantum systems, three fundamental factors must be considered~\cite{DAlessio_Kafri_Polkovnikov_Rigol_2016,Ghosh_Langlett_Hunter-Jones_Rodriguez-Nieva_2024}: \textit{(i)} \textit{Symmetries}, which determine the role of conserved quantities and how integrable systems reach equilibrium compared to non-integrable systems that lack such symmetries; \textit{(ii)} \textit{Initial conditions}, which shape the system's dynamics and outcomes based on the chosen starting state; and \textit{(iii)} \textit{Time scales}, representing the necessary evolution period to explore the system's phase space sufficiently and reach the quantum chaotic regime. In this work, we adopt a comprehensive approach that incorporates all of these key aspects, examining the quantum dynamics governed by local Hamiltonian evolution.

\begin{figure}[t!]
 		\centering 
   
\includegraphics[width=\columnwidth]{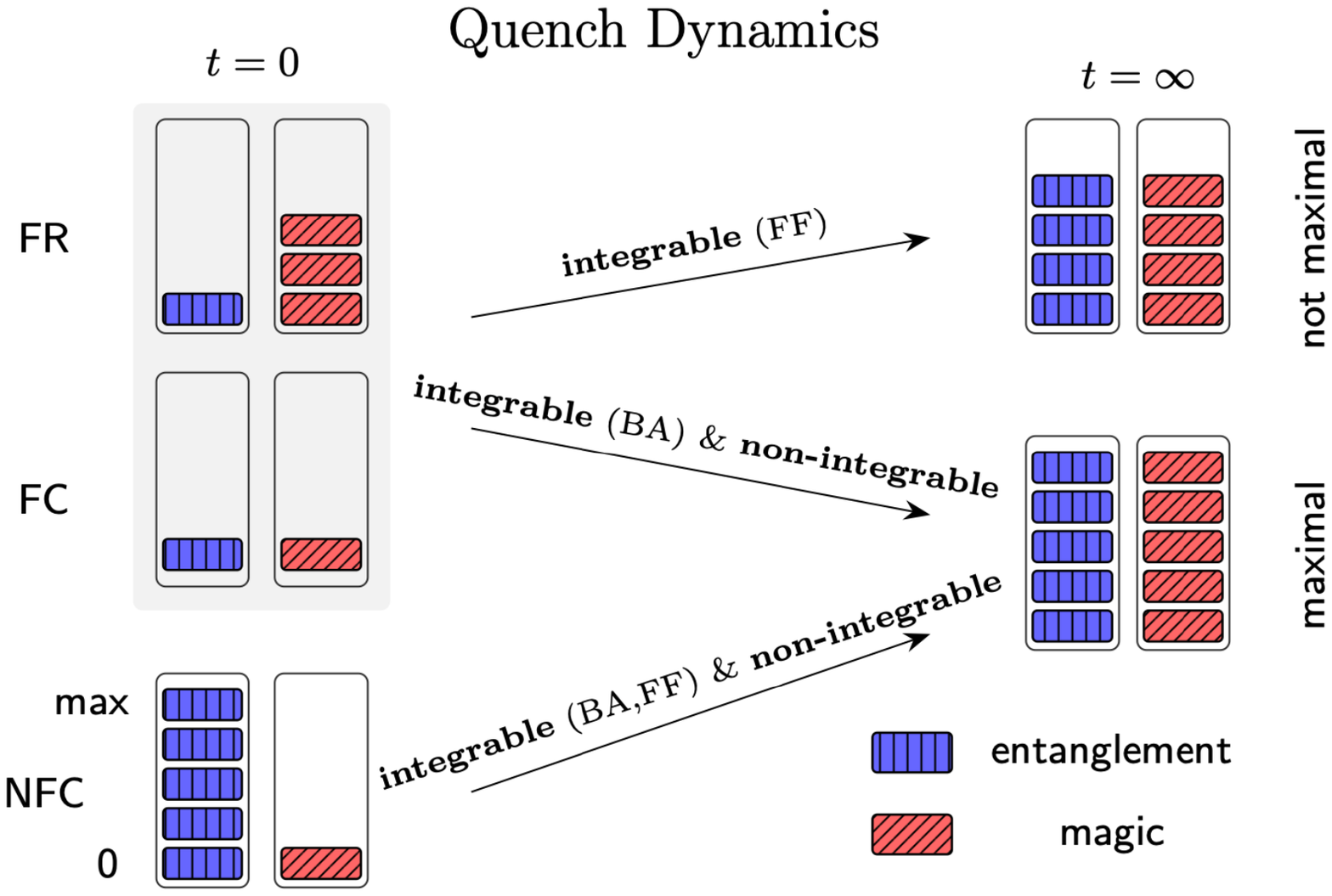}  
\caption{\justifying Schematic representation of the results for the Hamiltonian evolution of a global quantum quench protocol in the long-time limit. The initial states considered belong to three distinct ensembles: factorized random (FR), factorized Clifford (FC), and non-factorized Clifford (NFC) states. FR and FC ensembles display the same behavior and are grouped together by a common grey background color. A single colored rectangle (labeled ``0'' in the figure) indicates that the states possess zero amount of the particular resource. In contrast, five stacked rectangles denote the maximal value (``max'') expected from a typical random pure state. Notations: FF – free fermions; BA – Bethe Ansatz.}
\label{sketch}
 \end{figure}

Previous comprehensive studies on the topic of generic quench dynamics, such as Refs.~\cite{Rigol_2014I, Rigol_2014II, Rigol_2014III}, have not investigated the role of non-stabilizerness. More recent works, including Refs.~\cite{Goto_Nosaka_Nozaki_2021, Lami_Collura_2023}, have examined the SRE for integrable and non-integrable quench dynamics in Ising chains. However, these investigations were limited to short timescales and specific initial states, thereby restricting broader insights into the generic behavior of non-stabilizerness. On the other hand, approaches like the one in Ref.~\cite{Haug_Aolita_Kim_2024} randomized the Hamiltonian dynamics but did not specifically address the integrability of the dynamics nor the choice of initial states. In contrast, our work explores the time evolution of pure states that are not necessarily ground states of any local Hamiltonian but collectively addresses all three nuances raised above. Specifically, we consider the following ensembles of initial states:

\begin{enumerate}
\item Factorized Random (FR) states,
\item Factorized Clifford (FC) states, 
\item Non-Factorized Clifford (NFC) states.
\end{enumerate}

These initial conditions are randomized in a way to break any symmetry of the governing Hamiltonian. Moreover, they are chosen so that different starting points for the resources of magic and entanglement are met. FR states possess zero entanglement and extensive SRE, NFC states possess zero SRE and extensive entanglement, while FC states possess neither of the two resources. Notice that states belonging to the FR ensemble are completely unentangled (on any partition) but may exhibit maximal single-qubit magic. However, single-qubit magic represents only a portion of the total magic a state can host; a higher degree of magic can only arise in the presence of entanglement~\cite{Oliviero_Leone_Hamma_2022}.

Through the different quench protocols applied to states from these ensembles, we paint a detailed picture of the pathways leading to quantum chaos in Hamiltonian dynamics. Our findings demonstrate that breaking integrability results in universal behavior, which can be witnessed by both entanglement and magic. As illustrated in Fig.~\ref{sketch}, we provide a visual summary of the key results of this work. Notably, the concept of anti-flatness, as defined below, encapsulates both aspects—entanglement and magic—simultaneously, offering a unified perspective on these two crucial features. As our final contribution we supplement our results on entanglement and non-stabilizerness with out-of-time-order correlation (OTOC) function providing further evidence for the observed dynamical pathways to quantum chaos.

It has then become increasingly clear that it is the interplay between magic and entanglement that fosters quantum complex behavior. From quantum many-body theory to quantum thermodynamics, from black hole physics to nuclear physics, it is now apparent that one needs to investigate the deep connection between these two kinds of entropy, entanglement, and stabilizer~\cite{Robin_Savage_2024, Chernyshev_Robin_Savage_2024, Cepollaro_Chirco_Cuffaro_Esposito_Hamma_2024, Leone_Oliviero_Piemontese_True_Hamma_2022, Oliviero_Leone_Lloyd_Hamma_2024, Leone_Oliviero_Lloyd_Hamma_2024}. 
One probe into the joint actions of these resources is the anti-flatness of the reduced density operator~\cite{Tirrito_Tarabunga_Lami_Chanda_Leone_Oliviero_Dalmonte_Collura_Hamma_2024}. Antiflatness is indeed a measure of non-local magic and plays an important role also in the relationship between entanglement and back-reaction in AdS-CFT~\cite{Cao_Cheng_Hamma_Leone_Munizzi_Oliviero_2024, White_Cao_Swingle_2021}.

\section{Setting the stage}

We now outline the key components required to describe our approach. First, in Section~\ref{STS_InitialStates}, we explain the procedure for generating the initial state ensembles, with each state indexed by a subscript $m=1,2,..., M$, where $M$ denotes the ensemble's total size or cardinality. This section provides the necessary background on how these states are prepared for quench dynamics. Next, in Section~\ref{STS_QuantitiesOfInterest}, we define the entanglement and magic monotones that serve as the primary tools for quantifying the system's evolution. These measures are central to the analysis presented throughout the manuscript, enabling us to track and compare the behavior of different ensembles. A discussion on the importance of reduced density matrix eigenvalues and the associated property of entanglement and magic, the anti-flatness is given at the end of Section~\ref{STS_QuantitiesOfInterest}.

\subsection{Initial states}\label{STS_InitialStates}

\subsubsection{Factorized Random (FR) States}\label{RPSmodel}

We define the normalized pure FR state as a tensor product of single-qubit states:
\begin{align}
    \Big\vert \varphi_m^{\rm FR} \Big\rangle = \bigotimes_{j=1}^{N} \big\vert \varphi (\theta_j, \phi_j) \big\rangle, \label{randomproductstate}
\end{align}
where \( N \) represents the total number of qubits or spins. Each single-qubit state \( \vert \varphi (\theta_j, \phi_j) \rangle \) is given by:
\begin{align}
    \vert \varphi (\theta_j, \phi_j) \rangle = \cos\left( \frac{\theta_j}{2} \right) \vert 0 \rangle + e^{i \phi_j} \sin\left( \frac{\theta_j}{2} \right) \vert 1 \rangle,
\end{align}
with the angles \( \theta_j \in [0, \pi] \) and \( \phi_j \in [0, 2\pi) \) uniformly sampled for each qubit \( j = 1, 2, ..., N \) when numerically generating the state. By construction, the FR state is unentangled (i.e. trivial eigenvalues of the reduced density matrix) and exhibits only local non-stabilizerness (magic)~\cite{Oliviero_Leone_Hamma_2022, Odavic_Haug_Torre_Hamma_Franchini_Giampaolo_2023} due to single qubit rotations moving it away from the six possible stabilizer single qubit states $\{\vert 0 \rangle, \vert 1 \rangle, \vert + \rangle, \vert - \rangle, \vert i \rangle, \vert -i \rangle \}$~\cite{Nielsen_Chuang_2010}.

\subsubsection{Factorized Clifford (FC) States}\label{PCSmodel}

We generate a Factorized Clifford (FC) state using a quantum circuit. The initial state is the computational basis state \( \vert 0 \rangle^{\otimes N} \), which evolves as
\begin{align}
    \Big\vert \varphi_{m}^{\rm FC} \Big\rangle = U^{\rm RC} \big\vert 0 \big\rangle^{\otimes N}, 
\end{align}
under the action of a random Restricted Clifford (RC) circuit with depth \( N_{\rm layers} \). The RC circuit consists of multiple layers of gates, defined as
\begin{align}
    U^{\rm RC} = \prod\limits_{k = 1}^{N_{\rm layers}} U_{k} = \prod\limits_{k = 1}^{N_{\rm layers}} \prod\limits_{j}^{} U_{k}(j),
\end{align}
where the subscript \( k \) indicates a specific layer in the circuit. Each unitary \( U_{k} \) consists of \( N-1 \) single-qubit gates applied at uniformly sampled positions \( j \in \{1, 2, \dots, N \} \). These gates are randomly chosen from the set
\begin{align}
    U_{k}(j) \in \{ I_j, S_j, H_j \},
\end{align}
where \( I_j \) is the identity gate, \( S_j = \sqrt{Z}_{j} \) is the phase gate with \( \theta = \pi/2 \), and \( H_j =\frac{1}{\sqrt{2}}({X}_j+{Z}_j) \) is the Hadamard gate applied to qubit \( j \). ${X}, {Y}, {Z}$ refer to the Pauli matrices, which are used to express the unitaries compactly.

In practice, to ensure we obtain the typical representative of this set of states we set the number of Clifford layers to be \( N_{\rm layers} = 50 N^2 \)~\cite{Tirrito_Tarabunga_Lami_Chanda_Leone_Oliviero_Dalmonte_Collura_Hamma_2024,True_Hamma_2022}. The term "Restricted Clifford" refers to excluding two-qubit gates, such as the CNOT gate, which would introduce entanglement (see next subsection). As a result, the FC states do not alter (beyond equal weight) the amplitudes of the computational basis states and do not host entanglement, i.e. they are factorized.

\subsubsection{Non-Factorized Clifford (NFC) States}\label{ECSmodel}
\label{sec:NFC}
Compared to FC states, Non-Factorized Clifford (NFC) states are generated in a similar manner and with the same number of layers. However, the key distinction lies in the structure of the unitary operation \( U_{k} \), which now includes two-qubit gates. Specifically, \( N-1 \) two-qubit gates are applied at uniformly sampled qubit pairs \( (j, l) \in \{1, 2, \dots, N \} \) in each layer. These gates are randomly selected from the set:
\begin{align}
    {U}_{k} (j, l) &\in \{ {I}_j \otimes {S}_l, {S}_j \otimes {I}_l, {I}_j \otimes {H}_l, {H}_j \otimes {I}_l, \notag \\
    &\quad {\rm {CNOT}}_{j,l}, {\rm {CNOT}}_{l,j} \},
\end{align}
where \( {I}_j \), \( {S}_j \), and \( {H}_j \) represent the identity, phase, and Hadamard gates, respectively, acting on qubit \( j \). The gate \( {\rm {CNOT}}_{j,l} \) is the controlled-NOT (CNOT) gate acting on the \( l \)-th qubit, controlled by the \( j \)-th qubit, defined as
\begin{align}
    {\rm {CNOT}}_{j,l} = \exp\left[ i \frac{\pi}{4} (1-{X}_j)(1-{Z}_l) \right].
\end{align}
In contrast to FC states, NFC states generated through this process host extensive amounts of entanglement, yet remain magic-free. States generated in this way belong to the class of stabilizer states and the distribution of their eigenvalues is flat (see Section ~\ref{STS_QuantitiesOfInterest_AntiFlatness} and Fig.~\ref{fig:RDMeigenvalues}).

\subsection{Quantities of Interest}\label{STS_QuantitiesOfInterest}
\subsubsection{Rényi Entanglement Entropies}\label{STS_QuantitiesOfInterest_RE}

For pure states, quantum correlations are quantified by the Rényi entanglement entropies~\cite{Amico_Fazio_Osterloh_Vedral_2008, Nielsen_Chuang_2010}, which are defined as
\begin{align}
    S_{\alpha} (\mathbf{\rho}_{\rm R}) \equiv \! \frac{1}{1 - \alpha} \log{{\rm Tr} \left[ \rho^{\alpha}_{\rm R} \right]} = \! \dfrac{1}{1 - \alpha} \log \left( \sum\limits_{i = 1}^{2^{ R  }} \lambda_{i}^{\alpha} \right). \label{entropydef}
\end{align}
The Rényi entropy depends on the parameter \( \alpha \in [0, 1) \cup (1, \infty] \), and is computed from the reduced density matrix (RDM) \( \rho_{\rm R} \) on a subpartition \( R \), with the corresponding eigenvalues $\lambda_{i}$, where $\sum_{i} \lambda_{i} = 1$.  The RDM is constructed by tracing out the degrees of freedom in the complementary subpartition (denoted $R^{\rm c}$) as $\rho_{\rm R} \equiv {\rm Tr}_{R^{\rm c}} \vert \Psi \rangle \langle \Psi \vert $, where \( \vert \Psi \rangle \) represents the pure state of the entire system.

For $\alpha \to 1^{+}$, the Rényi entanglement entropy reduces to the von Neumann entanglement entropy
\begin{align}
    S_1(\rho_{\rm R}) &\equiv - {\rm Tr}[ \rho_{\rm R} \log(\rho_{\rm R}) ]  = - \sum_i \lambda_i \log(\lambda_i). \label{VNdef}
\end{align}
which captures the overall distribution of eigenvalues of $\lambda_{i}$ of the RDM, thus reflecting the average entanglement across a bipartition. On the other hand, $\alpha > 1$, the entropy begins to weigh the largest eigenvalues more heavily. An important property of Rényi entropies is their monotonicity, i.e. for two different values of \( \alpha \), if \( 1 < \alpha_1 \leq \alpha_2 \), the entropy satisfies the inequality
$S_{\alpha_1} (\psi) \geq S_{\alpha_2} (\psi)$~\cite{Muller-Lennert_Dupuis_Szehr_Fehr_Tomamichel_2013}.

\textit{R\'{e}nyi entanglement entropies of Haar states}. 
Haar random pure states are quantum states drawn uniformly from the Hilbert space according to the Haar measure. They represent the most generic states possible with no preferred structure or symmetries. Haar random pure states are crucial in quantum chaos as they exhibit maximal entanglement and complexity, mirroring the behavior of chaotic systems, where quantum dynamics leads to the rapid scrambling of information~\cite{Liu_Lloyd_Zhu_Zhu_2018,PhysRevLett.126.030601}. 

A lot is known analytically about these states. For example, in Refs.~\cite{Nakagawa_Watanabe_Fujita_Sugiura_2018, Liu_Lloyd_Zhu_Zhu_2018}, the explicit Page curves (volume law) for the scaling of the averaged Rényi entropy with positive integer $\alpha$ for Haar random pure states is given by (to leading order)
\begin{align}
S_{\alpha}& \left( \psi_{R}^{\rm Haar} \right)\! = \! \frac{1}{1 - \alpha} \log{ \left[2^{N - R(1 + \alpha) } \sum_{k=1}^{\alpha} H(\alpha, k) 2^{(2 R - N)k}\right]} \label{Page}
\end{align}
where the coefficients
\begin{align}
H(\alpha, k) = \frac{1}{\alpha} \binom{\alpha}{k} \binom{\alpha}{k-1},
\end{align}
are known as Narayana numbers~\cite{weisstein_narayana}. The remaining terms in the expression for the entanglement entropies scale as $\mathcal{O}(2^{-N})$, making them of vanishing contribution in the large $N$ limit.  Taking $\alpha \to 1^{+}$, one recovers the Page von Neumann entanglement entropy of these states~\cite{Page_1993}. Note that $ S_{\alpha} (\psi^{\rm Haar}) \equiv \mathbb{E}_{\rm Haar} \big[ S_{\alpha} \big]$, where $\mathbb{E}_{\rm Haar} [ \cdot ]$ is the average over an ensemble of states sampled according to the Haar measure.

For the half-chain case ($R = N/2$), different entropies to leading order in system sizes read
\begin{align}
    S_{1} \left( \psi_{N/2}^{\rm Haar} \right) &= \frac{N}{2} \log{(2)} - \dfrac{1}{2} \label{RE1} \\
    S_{2} \left( \psi_{N/2}^{\rm Haar} \right) &=  \frac{N}{2}\log{(2)} -  \log{(2)}, \label{RE2} \\ S_{3} \left( \psi_{N/2}^{\rm Haar} \right) &= \frac{N}{2} \log{(2)} - \dfrac{1}{2} \log{(5)}.
\end{align}
The monotonicity property of the Rényi entropies is correctly satisfied as $S_{1} (\psi_{N/2}^{\rm Haar}) > S_{2} (\psi_{N/2}^{\rm Haar}) > S_{3} (\psi_{N/2}^{\rm Haar})$.

\subsubsection{Stabilizer R\'{e}nyi Entropy}

The term ``magic''~\cite{Bravyi_Kitaev_2005} describes the property of quantum states that places them beyond the stabilizer formalism. It is widely regarded as a crucial component for universal quantum behavior, encapsulating what makes states inherently ``quantum''. To operationally quantify magic we employ the Stabilizer Rényi Entropy (SRE)~\cite{Leone_Oliviero_Hamma_2022}, defined as
\begin{align}
    \mathcal{M}_{\alpha} (\Psi) = \dfrac{1}{1 -\alpha} \log_{2}{ P_{\alpha} \left( \Psi \right) }, 
    \label{magicformula}
\end{align}
where $ \Psi = \vert \Psi \rangle \langle \Psi \vert $ density matrix  of the pure state $\vert \Psi \rangle$, and
\begin{align}
  P_{\alpha} (\Psi) = \dfrac{1}{d} \sum\limits_{P \in \mathcal{P}_{N}} \vert \langle \Psi \vert P \vert \Psi \rangle \vert^{2 \alpha},
\end{align}
is the stabilizer purity. For pure states, the SRE for $\alpha \ge 2$ is a good monotone from the point of view of resource theory~\cite{Leone_Oliviero_Hamma_2022}. However, they are not a strong monotone~\cite{Haug_Piroli_2023, Leone_Bittel_2024}. On the other hand, the linearized SREs 
\begin{align}
    \mathcal{M}_{\alpha}^{\rm lin} (\Psi) = 1 - P_{\alpha} (\Psi)
\end{align}
are strong monotones. In general, the stabilizer entropies are designed to quantify the spread
of a state in the basis of Pauli operators, and attempt to quantify the difference of a given state to those that are stabilizer states~\cite{Leone_Oliviero_Hamma_2022,Niroula_White_Wang_Johri_Zhu_Monroe_Noel_Gullans_2024}. The state $\vert \Psi \rangle \equiv \vert \Psi (N) \rangle $ is $N$-qubit state and $d = 2^{N}$ is the dimension of the Hilbert space, while $\alpha$ denotes the R\'{e}nyi index. With $\mathcal{P}_N$ we denote the set of Pauli strings built from the identity $I$ and Pauli operators $X, Y, Z$. The Hilbert space of $N$ spins grows exponentially as $2^{N}$ while the number of Pauli strings in the Pauli group grows as $4^{N}$ and the evaluation of SRE's implies a summation over an exponential number of expectation values over all possible up-to $N$ Pauli string correlation functions. Interesting properties of the SRE are:   
\begin{enumerate}[label=\roman*.]
    \item vanishing for pure stabilizer states, if $\Psi \in {\rm STAB}$ is $\mathcal{M}_{\alpha} (\Psi) = 0$ 
    \item invariant under the application unitary Cliffords $C$, as $\mathcal{M}_{\alpha} (C \Psi C^{\dagger}) = \mathcal{M}_{\alpha} (\Psi)$ 
    \item additive $\mathcal{M}_{\alpha} (\Psi \otimes \Phi) = \mathcal{M}_{\alpha} (\Psi) + \mathcal{M}_{\alpha} (\Phi)$,
    \item upper bounded as $\mathcal{M}_{\alpha} \leq \log_2{d}$, with a tighter bound in the case $\alpha \geq 2$ as $\mathcal{M}_{\alpha} \leq \log_{2} (d + 1) - 1$,
    \item lower bounds other well-known non-stabilizerness monotones such as the stabilizer nullity $\nu$ \cite{Beverland_Campbell_Howard_Kliuchnikov_2020}, min-relative entropy of magic~\cite{Liu_Winter_2022}, the robustness of magic~\cite{Howard_Campbell_2017}. 
\end{enumerate}
Compared to other non-stabilizerness monotones, the SRE is efficiently computable for small to intermediate-sized systems ($N < 12$ spins/qubits) without requiring a minimization procedure or approximations. Recent advancements have introduced efficient methods for evaluating pure states magic via the SRE using Matrix Product State (MPS) representations~\cite{Haug_Piroli_2023, Lami_Collura_2023, Lami_Haug_DeNardis_2024, Frau_Tarabunga_Collura_Dalmonte_Tirrito_2024, Tarabunga_Tirrito_Bañuls_Dalmonte_2024}. Although tensor network methods such as the MPS are potent, their efficiency is ultimately constrained by the amount of entanglement, i.e. the larger the bond dimension required to accurately approximate a quantum state, the less efficient the techniques to evaluate the SRE become. However, it is important to note that compared to entanglement the SRE is less sensitive to the growth of bond dimension~\cite{Frau_Tarabunga_Collura_Dalmonte_Tirrito_2024}. Note that there is a case where the phase transition of a many-body system in non-stabilizerness is not followed by one in entanglement in the thermodynamic limit~\cite{Catalano_Odavic_Torre_Hamma_Franchini_Giampaolo_2024}.

 \textit{Stabilizer R\'{e}nyi entropy of Haar states}. Recently, several authors have computed the magic of pure Haar random states using various techniques~\cite{Turkeshi2023,Haug_Aolita_Kim_2024, Turkeshi_Tirrito_Sierant_2025}, but in Ref.~\cite{Leone_Oliviero_Hamma_2022} the linearized SRE for these states was obtained and yields
\begin{align}
     \mathcal{M}_{2}^{\rm lin} (\psi^{\rm Haar}) = 1 - \dfrac{4}{d + 3}, 
\end{align}
and it simply follows in the large $N$ limit for qubits $d = 2^{N}$ that
\begin{align}
    \mathcal{M}_{2} (\psi^{\rm Haar}) = N - 2.
\end{align}
Note that we have used the notation $ \mathcal{M}_{\alpha} (\psi^{\rm Haar}) = \mathbb{E}_{\rm Haar} [\mathcal{M}_{\alpha}]$ to express the average over an ensemble of states sampled according to the Haar measure.

\begin{figure}[t!]
 		\centering \includegraphics[width=\columnwidth]{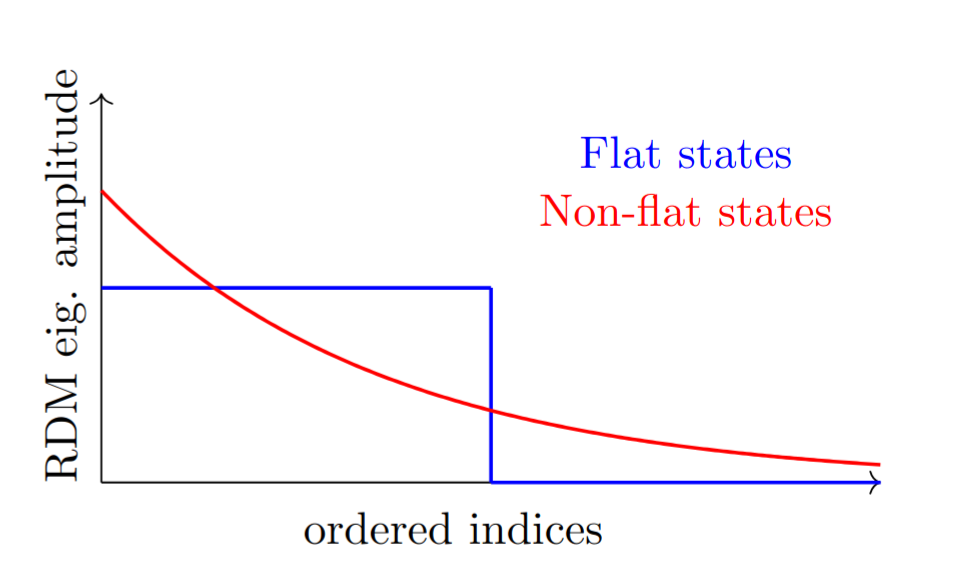}
 		\caption{\justifying Illustration of states with flat and non-flat distribution of RDM spectrum. 
      }
 \label{fig:RDMeigenvalues}
 \end{figure}

\subsubsection{Anti-flatness}\label{STS_QuantitiesOfInterest_AntiFlatness}

Lastly, we present some details about the recently introduced (anti)-flatness $\mathcal{F}$~\cite{Tirrito_Tarabunga_Lami_Chanda_Leone_Oliviero_Dalmonte_Collura_Hamma_2024}. It  is a function of the pure state's RDM $\mathcal{F}(\rho_{\rm R})$  across a bipartition $R \cup R^{\rm c}$ as,
\begin{align}
    \mathcal{F} (\rho_{R}) &:= {\rm Tr} \left[ \rho_{R}^{3} \right] - \left( {\rm Tr} \left[ \rho_{R}^2 \right] \right)^{2},\notag \\
    &= \sum\limits_{i=1}^{2^{R}} \lambda_{i}^3 - \left( \sum\limits_{i=1}^{2^{R}} \lambda_{i}^2 \right)^2  = \mbox{Var}_{\rho_R} (\rho_R)
    \label{F1def} 
\end{align}
and is a property of the entanglement spectrum. More specifically, the anti-flatness measures how much the distribution of the RDM eigenvalues deviates from a flat (or uniform) distribution. Here ${\rm Pur}(\rho_{R}) = {\rm Tr} [\rho^{2}]$ defines the purity. The spectrum is flat when the spectrum $\lambda_{i} = 1/\chi$ for some integer $1 \leq \chi \leq \min(d_R, d_{R^{\rm c}})$. For $\mathcal{F}_A (\psi) \neq 0$, the state contain magic. Interestingly, stabilizer states are known to be a class of states that admit a flat distribution in the entanglement spectrum~\cite{Hamma_Ionicioiu_Zanardi_2005}. Fig.~\ref{fig:RDMeigenvalues} gives a simple illustration of flat and non-flat states.

In Ref.~\cite{Tirrito_Tarabunga_Lami_Chanda_Leone_Oliviero_Dalmonte_Collura_Hamma_2024} an important result regarding the link between anti-flatness and magic has been provided. More specifically  
\begin{align}
    \left\langle \mathcal{F} (\rho_{R}) \right\rangle_{\rm C} = c (d, d_{R}) M^{\rm lin}_{2} (\vert \psi \rangle).
\end{align}
Here $\langle \cdot \rangle_{\rm C}$ denotes the average over the Clifford orbit, while $c(d,d_{R})$ is a proportionality constant that depends on the total and subsystem size $d$ and $d_{R}$, respectively. The Clifford orbit is a necessary device for any state that is not volume law entangled for the link between the linearized SRE and the anti-flatness to exist. {As we illustrate in this work, in global quantum quench dynamics the time-evolved states in the long-time limit naturally develop extensive (volume-law) entanglement making the anti-flatness a suitable tool to differentiate between the different pathways towards equilibrium in the considered settings. }

\textit{Anti-flatness of Haar states}. Looking back at Eq.~\eqref{Page} we observe it contains the closed-form expressions for the general RDM. We can easily read off $\alpha = 2,3$ and obtain for the half-chain partition $R = N/2$ the following 
\begin{align}
    {\rm Tr} \left[ \left( \psi_{N/2}^{\rm Haar} \right)^{2} \right] &= 2^{1 - N/2},  \label{moment2} \\
    {\rm Tr} \left[ \left( \psi_{N/2}^{\rm Haar} \right)^{3} \right] &= 5 \cdot 2^{-N}. \label{moment3}
\end{align}
Simply taking the difference we obtain
\begin{align}
    \mathcal{F} (\psi^{\rm Haar}_{N/2}) \sim 2^{-N},
\end{align}
which tells us that for Haar random pure states, the anti-flatness decays exponentially, and resolving such a small number in a large $N$ limit is challenging for typical states of the Hilbert space. Again we use  $ \mathcal{F} (\psi^{\rm Haar}) = \mathbb{E}_{\rm Haar} [\mathcal{F}]$ to express the average over an ensemble of states sampled according to the Haar measure.

Due to the requirement of evaluating an exponentially small number to obtain the anti-flatness of a typical state, we resort to a different quantity, the ``logarithmic anti-flatness''. Which we define as the difference between the R\'{e}nyi entropies reading
\begin{align}
    \mathrm{F} (\rho_{R}) &:= 2 \left(  S_{2} (\rho_{R}) - S_{3} (\rho_{R}) \right), \notag \\
    &= \log{\left( \sum\limits_{i=1}^{2^{R}} \lambda_{i}^{3} \right)} - \log{\left( \sum\limits_{i=1}^{2^{R}} \lambda_{i}^2 \right)^2}.
\end{align} 
Due to the monotonicity of R\'{e}nyi entropies, logarithmic anti-flatness is always a positive number~\cite{Muller-Lennert_Dupuis_Szehr_Fehr_Tomamichel_2013}. To quantify the logarithmic anti-flatness of the RDM eigenvalues distribution any difference between R\'{e}nyi entropies would do, but here we simply focus on the smallest ($\alpha = 2$ and $\alpha = 3$) indices choice. 

\textit{Logarithmic anti-flatness of Haar states}. Immediately, we can evaluate the quantity for this class of states using the results from Sect.~\ref{STS_QuantitiesOfInterest_RE} to be
\begin{align}
   \mathrm{F} \left( \psi^{\rm Haar}_{N/2}\right) = \log{\left( \dfrac{5}{4} \right)}.
\end{align}
This quantity is a constant compared to the first definition for the typical states and does not require computing an exponentially small number as the number of qubits is considered. This makes it more suitable for evaluation in numerical simulations with many qubits/spins.

\section{Quench Dynamics}
\subsection{Description}
\label{sec:description}
We perform a global quench protocol~\cite{Calabrese_Cardy_2005, Alba_Calabrese_2018, Calabrese_2020, Torre_Odavic_Fromholz_Giampaolo_Franchini_2024}. Specifically, the quantum state of the system evolves as:
\begin{align}
\left|\psi_0 \right\rangle \mapsto\left|\psi_t \right\rangle = e^{-i \mathcal{H} t}\left|\psi_0 \right\rangle,
\end{align}
where $t$ denotes the time and $\mathcal{H}$ is the quench Hamiltonian. The time evolution is unitary, preserving the norm of the state vector. The initial states $\left|\psi_0 \right\rangle$ are selected from the ensembles described in Sect.~\ref{STS_InitialStates}, which include various factorized and non-factorized ensembles. By investigating the response of different ensembles to quench dynamics, we aim to assess how integrable and non-integrable behavior manifests across various many-body Hamiltonians.

To efficiently simulate the time evolution, we employ the Krylov subspace method, a powerful technique for reducing the computational complexity of time evolution in intermediate-to-large quantum systems. This method constructs a lower-dimensional subspace that captures the essential dynamics of the full Hilbert space, allowing us to bypass the need for full diagonalization of the Hamiltonian. The key advantage of this approach lies in its scalability and applicability to large system sizes, making it particularly suited for studying long-time dynamics in quantum many-body models~\cite{Luitz_Laflorencie_Alet_2016, Nauts_Wyatt_1983}. In practice, we keep the relative truncation errors of the time-evolved states under control and at around $\sim 10^{-12}$. The truncation errors do not accumulate over the long-time evolution with a simple choice of an appropriate time-step, determined as a function of the final target time. In the following, we specify those choices in the captions of the figures we present. Naturally, the efficiency of the Krylov subspace method is bounded by the ability to store the full eigenstate.

For the Hamiltonians generating the time evolution, we focus on the following one-dimensional spin-1/2 models, chosen for their relevance in the study of both integrable and non-integrable quantum systems:

\textit{(i) Transverse-field Ising model (TFIM) + longitudinal (L) field}
\begin{align}
\mathcal{H}^{\rm TFIM + L} \!=\! - J \sum\limits_{j=1}^{N} X_{j} X_{j+1}\! -\! h_{z} \sum\limits_{j=1}^{N} Z_{j}
  \!-\! h_{x} \sum\limits_{j=1}^{N} X_{j}. \label{TFIM_Hamiltonian}    
\end{align}
In the absence of the longitudinal field ($h_x = 0$), TFIM can be solved exactly through a mapping to free fermions using the Jordan-Wigner transformation~\cite{Mbeng_Russomanno_Santoro_2024, Lieb_Schultz_Mattis_1961, Barouch_McCoy_1971, Franchini_2017}. This transformation maps the spin operators to fermionic creation and annihilation operators, effectively reducing the interacting spin system to a system of non-interacting fermions.

During time evolution, we set the transverse field such that the system is away from criticality, fixing $J = 1, h_{z} = 1.5$. By introducing a small-to-intermediate longitudinal field $h_x \neq 0$, we break the integrability of the model, allowing us to explore how the system's dynamics deviate from the integrable behavior.

\textit{(ii) XXZ chain + next-to-nearest-neighbor (NNN) coupling}
\begin{align}
\mathcal{H}^{\rm XXZ + NNN} &=  \sum_{j=1}^N\Big[ X_j X_{j+1} + Y_{j} Y_{j+1} + \Delta Z_{j} Z_{j+1} \notag \\
&+\alpha\left(X_{j} X_{j+2} + Y_{j} Y_{j+2} + \Delta Z_{j} Z_{j+2} \right) \Big].\label{XXZ_Hamiltonian}
\end{align}
The XXZ model, with additional next-to-nearest-neighbor (NNN) interactions, admits a rich phase diagram encompassing both integrable and non-integrable regimes. We differentiate three distinct regimes based on the parameters $\Delta$ (anisotropy) and $\alpha$ (strength of the NNN coupling):
\begin{enumerate}
    \item Integrable via the \textit{Free Fermion technique}~\cite{Mbeng_Russomanno_Santoro_2024, Lieb_Schultz_Mattis_1961, Barouch_McCoy_1971, Franchini_2017} 
    \begin{align}
        \Delta = 0,\quad {\rm and} \quad \alpha = 0,
    \end{align}
    In this regime, the XXZ chain maps to free fermions, similar to the TFIM. The lack of interactions between the fermions results in a highly tractable, integrable system.

    \item Integrable via the \textit{Bethe Ansatz}~\cite{Franchini_2017} 
    \begin{align}
        \Delta \neq 0, \quad {\rm and} \quad \alpha = 0,
    \end{align}
    For non-zero anisotropy $\Delta$, the system remains integrable but cannot be mapped to free fermions. Instead, the Bethe Ansatz provides an exact solution, relying on a set of coupled algebraic equations that describe the system's eigenstates. 
    \item Non-integrable 
    \begin{align}
        \forall \Delta  \quad {\rm and} \quad \alpha \neq 0.
    \end{align}
    In this regime, the addition of NNN interactions ($\alpha \neq 0$) breaks the integrability of the model, leading to chaotic dynamics and complex behavior.
\end{enumerate}
 We impose periodic boundary conditions to eliminate edge effects in both models.
 \begin{figure*}[t!]
 		\centering \includegraphics[width=18cm]{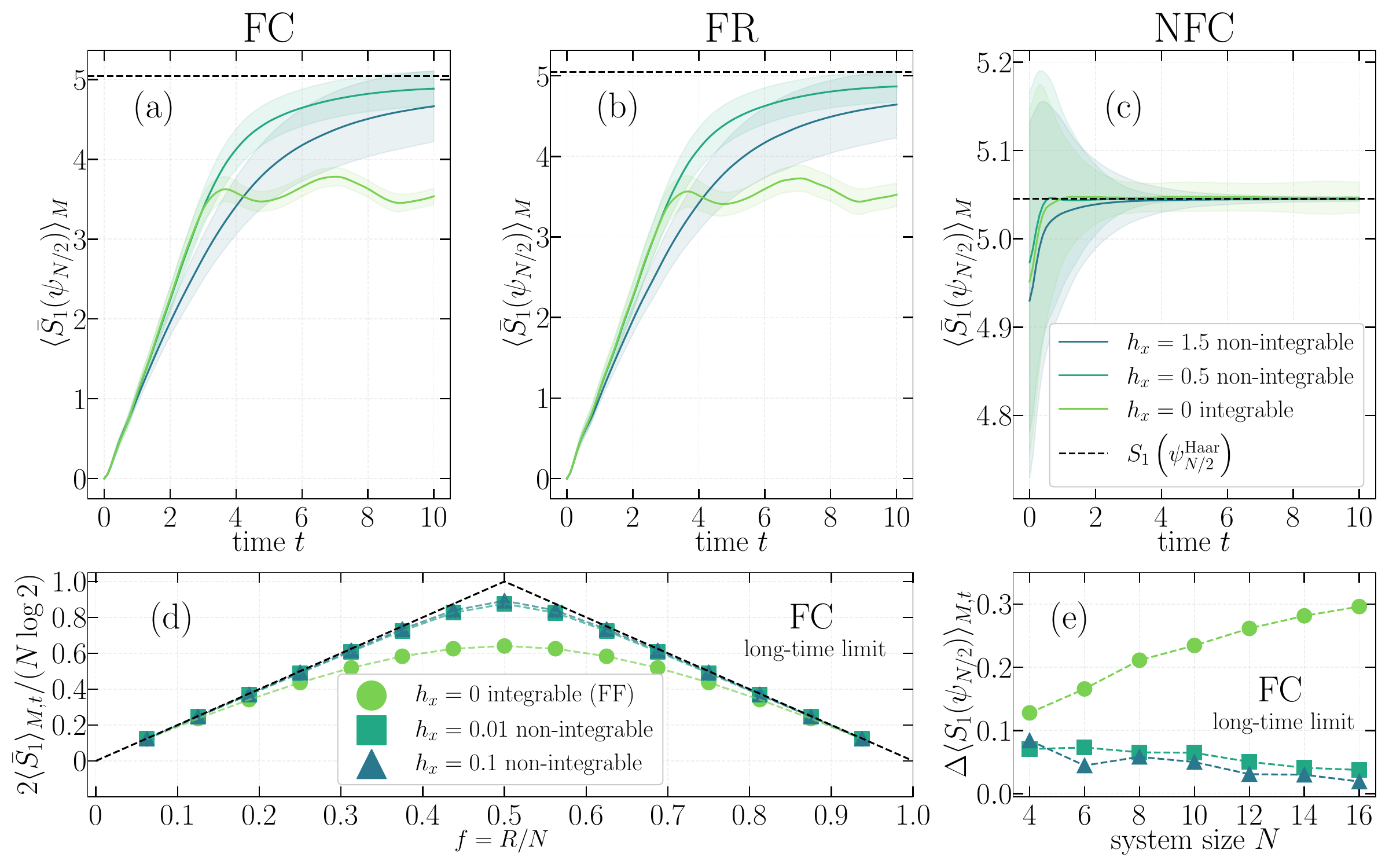}
 		\caption{ \justifying \textit{Panels (a-c)}: Half-chain von Neumann entanglement entropy Eq.~\ref{VNdef} for short-time dynamics generated by TFIM-L Hamiltonian and different initial state ensembles (see Sect.~\ref{STS_InitialStates}): Factorized Clifford (FC), Factorized Random (FR), and Non-Factorized Clifford (NFC). Parameters choices: $N = 16$, $M = 100$ realizations, and $h_{z} = 1.5$ and the time-step $\Delta t = 0.1$. The legends are shared between the panels (a-c). The shaded areas represent the standard deviation from the sample mean across $M$ realizations of the initial states. \textit{Panel (d)}: Average entanglement entropy for different subsystem partition sizes in the long-time limit where the initial states are sampled from the FC ensemble. Parameters choices: $N = 16$, $M = 50$ realizations, $\Delta t = 2$,   $t^{\rm final} = 10^{4}$. \textit{Panel} (e): Relative difference as defined by Eq.~\ref{relative_difference} of the von Neumann entanglement entropy for parameter choices as in panel (d) but for increasing system sizes. Non-integrable dynamics in the long-time limit lead to Haar random states, while integrable do not. The legends are shared between panels (d-e).}
 \label{fig:short1}
 \end{figure*}
In the following, we examine the Hamiltonian dynamics of these models, focusing on how integrability influences the system's entanglement, magic, and anti-flatness in the out-of-equilibrium regime. We also explore how these models compare to quantum chaos (as captured by by pure state sampled according to the Haar measure) predictions in their long-time behavior, emphasizing the role of initial state ensembles in shaping the dynamical response of the system.

\subsection{Results} 
In Figs.~\ref{fig:short1} (a-c), we show the short-time ($t^{\rm final} = 10$) dynamics of the TFIM+L Hamiltonian, focusing on the von Neumann entanglement entropy, which is defined in Eq.~\eqref{VNdef} and corresponds to the limit $\alpha \to 1^{+}$ of the R\'{e}nyi entanglement entropies. The entanglement entropy is averaged over contiguous (connected) subpartitions of the system and computed as
\begin{align}
    \overline{S_{\alpha} }(\psi_{R})   = \frac{1}{N} \sum_{i=1}^{N} S_{\alpha} (\psi_{R_{i}}), \label{S2}
\end{align}
where $N$ denotes the number of spins/qubits and the number of possible bipartitions $R_{i}$ (of size $R$) of the system, represented here by the overline notation. The initial states for the quench protocol are selected from the ensembles described in Sect.~\ref{STS_InitialStates}, and each observable is further averaged as
\begin{align}
    \langle \overline{S_{\alpha}}(\psi_{R}) \rangle_{M} = \dfrac{1}{M} \sum\limits_{m = 1}^{M} \overline{S_{\alpha}}_m (\psi_{R}).
\end{align}
The shaded areas represent the standard deviation from the sample mean across $M$ realizations of the initial states.

For both the FC and FR ensembles (panels (a) and (b), respectively), the initial states are unentangled, and their R\'{e}nyi entropy is zero by construction. As the states evolve after the sudden quench, the entanglement grows linearly at short times for both ensembles~\cite{Calabrese_Cardy_2005}. After this rapid initial increase, the entanglement dynamics of these initial ensembles follow similar qualitative trends, yet the integrable and non-integrable cases remain distinguishable. Notably, even at short timescales, the entanglement in integrable dynamics is significantly smaller, on average, compared to its non-integrable counterpart.

In Fig.~\ref{fig:short1} (d), we extend this analysis to much longer times, up to $t^{\rm final} = 10^4$, compared to the results shown in the upper panels. While the data is taken for initial states from the FC ensemble, we observe similar behavior for the FR ensemble (figure omitted as effectively conveys the same observation), underscoring that the distinction between integrable and non-integrable dynamics is inherent to the Hamiltonian dynamics. We average over multiple final time steps for this figure as
\begin{align}
     \langle \overline{S_{\alpha}}(\psi_{R}) \rangle_{M,t} = \dfrac{1}{T} \sum\limits_{j = 1}^{T} \langle \overline{S_{\alpha}}(\psi_{R}) \rangle_{M} (t_{j})
\end{align}
where $t_{j} = (5 \cdot 10^{3} + 1) \Delta t - j \Delta t$, $\Delta t = 2$, and $T = 100$.

Additionally, we plotted the normalized von Neumann entanglement entropy, focusing solely on the subsystem-to-system size ratio. The dashed black line represents the Page value $S_{1} (\psi^{\rm Haar}_{R}) = R \log(2)$ for random pure states in the $N \to \infty$ limit rescaled to
\begin{align}
    \dfrac{2 S_{1} \left( \psi_{R}^{\rm Haar} \right)}{N \log (2)} = 2 f, \quad {\rm for} \quad f \in[0,1/2].
\end{align}
Here, $f = R /N$ is the subsystem-to-system ratio, and for $f \in [1/2,1]$, the Page value is obtained by the transformation $f \to 1 - f$.
\begin{figure*}[t!]
 		\centering
 		\includegraphics[width=18cm]{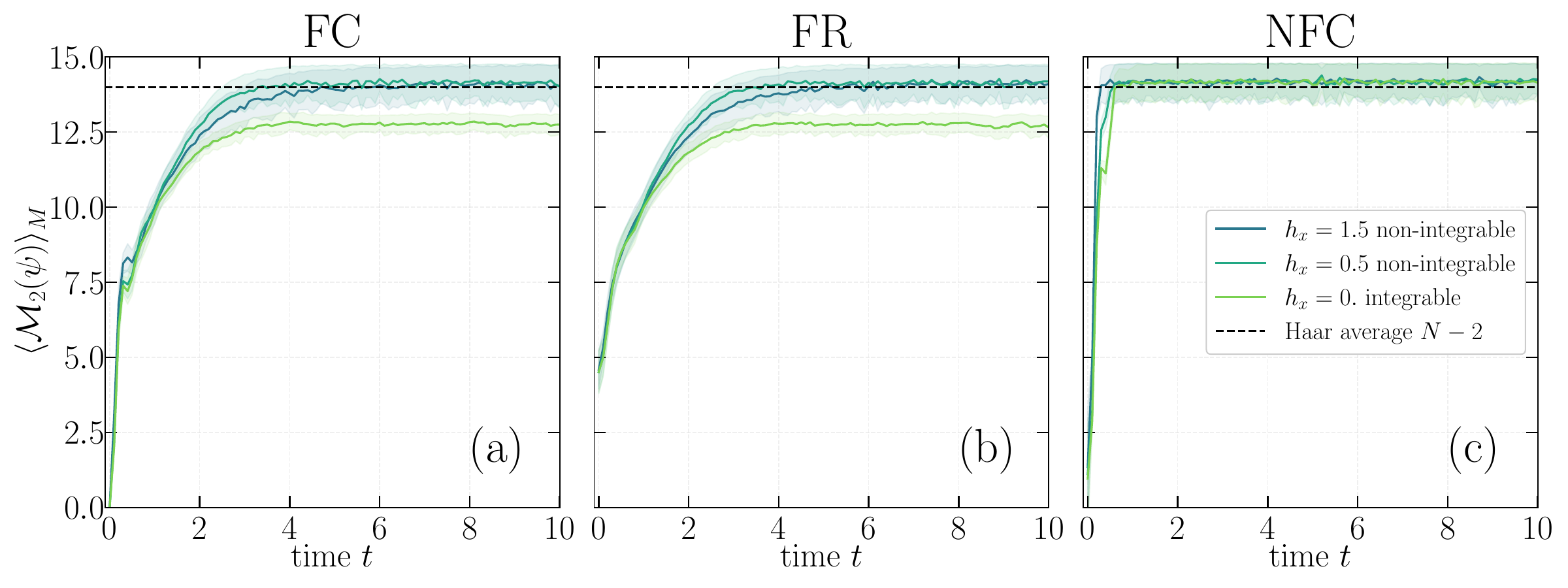}
 		\caption{\justifying \textit{Panels (a-c)}: Stabilizer R\'{e}nyi entropy $\mathcal{M}_{2}$ measuring the amount of non-stabilizerness (magic) for short-time $t \in [ 0, 10]$ dynamics generated by TFIM+L Hamiltonian and different initial state ensembles (see Sect.~\ref{STS_InitialStates}): Factorized Clifford (FC), Factorized Random (FR), and Non-Factorized Clifford (NFC). Parameters choices: $N = 16$, $h_{z} = 1.5$, and the time-step $\Delta t = 0.1$ with $ M = 100$ realizations. Entanglement entropy for the same parameters is given in Fig.~\ref{fig:short1}. The shaded areas represent the standard deviation from the sample mean across $M$ realizations of the initial states.
 		}
 \label{fig:short2}
\end{figure*}

The behavior of the long-time averaged states subjected to integrable dynamics is similar to the behavior of the typical states of random quadratic Hamiltonians~\cite{Vidmar_Hackl_Bianchi_Rigol_2017, Nakagawa_Watanabe_Fujita_Sugiura_2018,  Rigol_Vidmar_2020,  Bianchi_Hackl_Kieburg_Rigol_Vidmar_2022}. Both the integrable ($h_{x} = 0$) and non-integrable ($h_{x} \neq 0$) dynamics display a volume law, but with the difference that in the integrable case, the volume-law coefficient depends on the subsystem fraction~\cite{Rigol_Vidmar_2020}. More specifically, the leading order behavior of entanglement entropy is still present. Still, its magnitude additionally depends on the $f $, making it smaller than the maximal value for $f > 0$~\footnote{In Ref.~\cite{Rigol_Vidmar_2020} the explicit expression has been derived for random quadratic Hamiltonian eigestates and reads $S (\psi_{R}^{\rm quadratic})\left( f \right)=\left[1-\frac{1+f^{-1}(1-f) \ln (1-f)}{\ln 2}\right] R \ln 2$. However, obtaining a direct comparison (overlap of our results with this expression) is hard to make due to finite-size numerics and other details such as the filling factor on top of the fact that here we consider a particular long-time limit of the quench.}. Random Gaussian states that are eigenstates of the quadratic Hamiltonians typically lead to such entanglement entropy scaling. This type of state also captures eigenstates of quadratic Sachdev-Ye-Kitaev (SYK2) model~\cite{Rigol_Vidmar_2020}. We note that the results for the typical states of the random quadratic Hamiltonians are obtained for ensemble-averaged (typical) middle-of-the-spectrum states~\footnote{ In many-body quantum systems, especially those described by local Hamiltonians, the energy spectrum typically ranges from a minimum. i.e. the ground state to a maximum or highest excited state. Middle-of-the-spectrum states are near the center of the system's energy spectrum and inbetween the ground state and the highest excited state. Such states are often considered typical in the sense that they represent the majority of possible states in large quantum systems}, while here we are performing a long-time average, but also an ensemble average over the initial states. 

Performing quench dynamics starting from particular initial states, such as the ground states of the Ising chain, may not immediately lead to dynamics that saturate entanglement to a volume law behavior at short to intermediate time scales~\cite{Alba_Calabrese_2018, Kormos_Collura_Takacs_Calabrese_2017}. An example of such a quench protocol is presented in Fig.~(5) of Ref.~\cite{Lami_Collura_2023}, where the entanglement generated via non-integrable quench protocol deviates significantly from volume law, i.e. universal behavior. As the entanglement and magic are closely related resources this induces also non-universal behavior in magic as well~\cite{Oliviero_Leone_Hamma_2022}.  We address the challenge of accessing the behavior of Hamiltonian dynamics by randomizing the initial state, as previously discussed, which allows us to reach the equilibrium regime within computationally feasible simulation run times.

To exemplify the sensitivity of the dynamics to the integrability-breaking parameter, in Fig.~\ref{fig:short1} (d) and (e), we deliberately chose very small (relative to the remaining scales of the model) values for the integrability-breaking parameter, $h_{x} \ll 1$, i.e., $h_{x} \in [0.01, 0.1]$ (represented by the teal square and blue triangle, respectively). Despite the small values, the difference in bipartite entanglement is substantial. It is important to note that such small parameters result in dynamics that take much longer to reach their steady-state regime compared to the parameter choices and short-time evolution shown in the upper panels of Fig.~\ref{fig:short1} (a-c).

While the non-integrable dynamics approach the Haar state expectation, represented by the black dashed line and Eq.~\eqref{RE1}, the integrable dynamics oscillate around a value significantly lower than this expectation. To quantify this difference more precisely, in the final panel Fig.~\ref{fig:short1} (e), we compute the relative difference defined as
\begin{align}
    \Delta A = \frac{\vert A - A^{\rm Haar} \vert}{A^{\rm Haar}}, \label{relative_difference}
\end{align}
for the von Neumann entanglement entropy (where $A = S_{1}$). This shows that the finite-size numerical results for the non-integrable quench dynamics quickly converge to the thermodynamic limit of the Haar random state value and universality. In contrast, the integrable dynamics exhibit a much larger relative difference, highlighting their distinct behavior.

It is important to note that for the NFC ensemble of initial states, as already hinted in Fig.~\ref{fig:short1} (c), both integrable and non-integrable dynamics ultimately lead to universal behavior. Specifically, if we were to present the results under the same conditions as in panel (e), both integrable and non-integrable data for the NFC ensemble initial states would converge to zero as the system size increases. This is particularly intriguing, as it reveals the different pathways that dynamics can take depending on the specifics of the initial conditions.

Another important detail is also evidenced by our short-time dynamics results plotted in Fig.~\ref{fig:short1} (a-c). It is the difference between the fluctuations displayed by the averaged entanglement entropy between the integrable and non-integrable dynamics for all the considered ensembles~\cite{Hamma_Santra_Zanardi_2012, Venuti_Zanardi_2013}. The integrable dynamics lead to much larger fluctuations in time than the non-integrable one for all considered initial state ensembles, making this feature independent of the initial state. {A similar observation can be made in the results shown in Ref.~\cite{Lami_Collura_2023} when the initial states are eigenstates of the Hamiltonian.}

In quantum circuit dynamics, it has been shown that the fluctuations in subsystem purity~\footnote{Subsystem purity is defined as ${\rm Pur} (\rho_{R}):= {\rm Tr} [\rho_{R}^2]$, which relates to entanglement through its logarithm; see Eq.~\ref{entropydef}.} of output states generated by gates sampled from the full unitary group (concerning the Haar measure) and the Clifford group exhibit notable differences~\cite{Leone_Oliviero_Hamma_2022}. Specifically, for unitary group sampling, these fluctuations scale as $\mathcal{O} (d^{-2})$, whereas for Clifford sampling, they scale as $\mathcal{O} (d^{-a})$. The value of $a$ depends on the input state: $a = 1$ when the input is the product state $\vert 0 \rangle \langle 0 \vert^{\otimes N}$, and $a = \log_2{(5)} - 1$ for any other stabilizer input state (see Sect.~(4) of Ref.~\cite{Leone_Oliviero_Zhou_Hamma_2021}). Regardless of the initial state, the distinction between non-universal (Clifford) and universal (Haar) sampling is stark, making the output states distinguishable. To induce a transition towards universality, it has been proposed that the insertion of $\mathcal{O}(N)$ T-gates can achieve this effect~\cite{Leone_Oliviero_Zhou_Hamma_2021}. 

Given these findings, one may wonder whether the differences observed between integrable and non-integrable Hamiltonian dynamics might parallel those seen in $t$-doped quantum circuits. Although circuit dynamics evolve through discrete gates and Hamiltonian dynamics unfold continuously over time, making them fundamentally distinct, the existence of a possible analogy is interesting.

The purity fluctuations observed in quantum circuit dynamics served as the initial motivation for developing the SRE~\cite{Leone_Oliviero_Hamma_2022}, aimed at quantifying the non-stabilizerness resources introduced by T-gates, which drive the dynamics towards universality. By a similar token, we apply the SRE to investigate  the Hamiltonian dynamics. We must immediately underline that the non-integrable Hamiltonian eigenstates possess an extensive amount of non-stabilizerness as expected from many-body systems~\cite{Oliviero_Leone_Hamma_2022, Rattacaso_Leone_Oliviero_Hamma_2023, Liu_Winter_2022}, so any direct comparison is expected to fail. 

In Fig.~\ref{fig:short2}, we present the results of the short-time dynamics of the SRE, analogous to the analysis previously conducted for the same time data in Fig.~\ref{fig:short1} and for the entanglement entropy. Similar to the behavior of entanglement entropy, initial states from both the FC and FR ensembles display distinct differences between integrable and non-integrable dynamics. Specifically, integrable dynamics oscillate around a lower average value than what would be expected if the dynamics were universal. A key distinction between the FC and FR ensemble states is the amount of magic present at the initial time. Nonetheless, despite this difference, the two ensembles exhibit qualitatively similar dynamical behavior.

In contrast, the Hamiltonian dynamics of initial states sampled from the NFC ensemble equilibrate to values consistent with those of the Haar ensemble, regardless of whether the dynamics are integrable or non-integrable. This behavior mirrors what is observed for entanglement entropy. The study of NFC ensemble states highlights the crucial role of entanglement in the initial states for determining both the quench time evolution and the system's average behavior. In this case, entanglement and magic reach universality simultaneously and to the same level.

To compute the SRE, we employ the Pauli Perfect Sampling algorithm introduced in Ref.~\cite{Lami_Collura_2023}. Starting from the time-dependent state vector obtained via the Krylov subspace method, we convert it into a MPS representation and sample $10^{4}$ Pauli strings. This approach bounds the magic estimation error to approximately $10^{-2}$ when no truncation is applied to the bond dimension. However, in this work, we approximate the bond dimension by halving its maximum value, which introduces an estimated absolute error of $10^{-1}$ in the reported SRE results. We find that for our current work, this precision is sufficient to highlight the main messages.  An alternative method for efficiently computing the SRE directly from the MPS representation is described in Ref.~\cite{Tarabunga_Tirrito_Bañuls_Dalmonte_2024}. For comparison, sampling techniques that operate directly on the state vector, such as those in Refs.~\cite{Tarabunga_Tirrito_Chanda_Dalmonte_2023, Turkeshi2023}, generally exhibit significantly larger errors than the Pauli Perfect Sampling algorithm. 
One more comment must be added on Fig.~\ref{fig:short1}. Here, in panel (c), the averaged 2-SRE at the initial time exhibits a larger error than the theoretical prediction. The reason for this error is to be found in the bond dimension's truncation, which at $t=0.0$ involves a stabilizer state that is also typically highly entangled. The effect of the truncation is therefore to perturb the stabilizer state, thus injecting some magic.

\begin{figure}[t!]
 		\centering \includegraphics[width=\columnwidth]{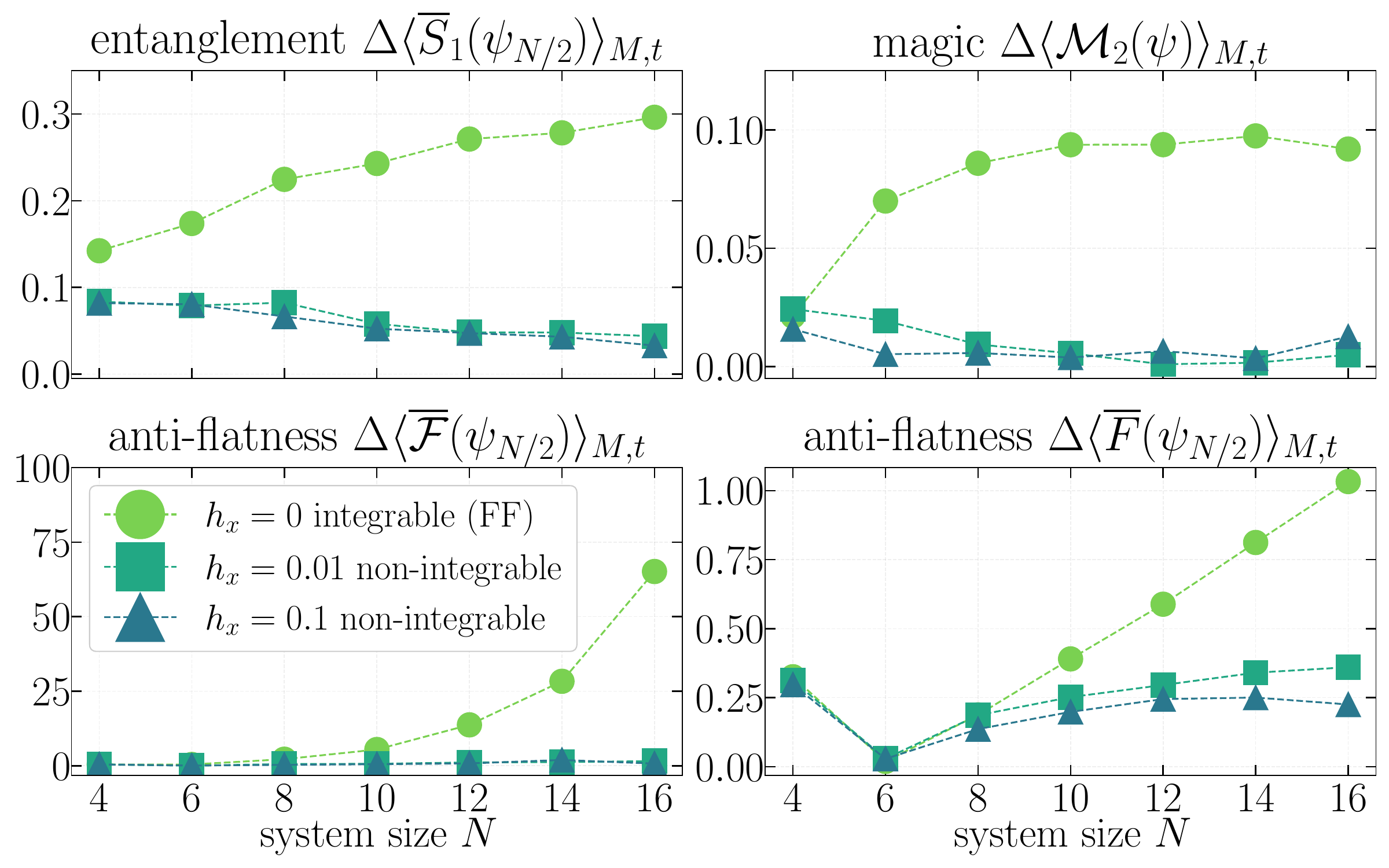}
 		\caption{ \justifying
    Ensemble averaged relative difference defined for different quantities in the case of the long-time quench dynamics generated by the TFIM+L Hamiltonian. Parameters choices: $M = 50$, $\Delta t = 2$,   $t^{\rm final} = 10^{4}$ (long-time limit), and FR as the initial state ensemble.
      }
 \label{fig:TFIM}
 \end{figure}

In Fig.~\ref{fig:TFIM}, we present results for the late-time quench dynamics using initial states sampled from the FR ensemble, with the evolution governed by the TFIM+L Hamiltonian. In the long-time limit, both entanglement and magic (upper row) reveal a clear distinction between integrable and non-integrable dynamics. Similar behavior is observed when the initial states are sampled from the FC ensemble. However, for initial states drawn from the NFC ensemble, this distinction between integrable and non-integrable quenches disappears, which can already be inferred from Fig.~\ref{fig:short1} and \ref{fig:short2} (c). These findings highlight the critical role of the initial state's entanglement in shaping the global quench dynamics, suggesting that the pathways to long-time behavior and universality strongly depend on the entanglement properties of the initial state.

Moreover, to simultaneously quantify entanglement and magic, we utilize the recently introduced (anti)-flatness measure, $\mathcal{F}(\rho_{\rm R})$. As shown in Fig.~\ref{fig:RDMeigenvalues}, a flat distribution of reduced density matrix (RDM) eigenvalues is a hallmark of stabilizer states. For instance, the eigenstates of the toric code, which are stabilizer states, exhibit a flat RDM eigenvalue distribution~\cite{Hamma_Ionicioiu_Zanardi_2005}. Any deviation from this flat spectrum provides a measure of the magic content in the total state~\cite{Tirrito_Tarabunga_Lami_Chanda_Leone_Oliviero_Dalmonte_Collura_Hamma_2024}. 

It is important to note that non-entangled states - those across a bipartition with only one non-zero RDM eigenvalue - can still possess magic. However, this magic arises purely from local single-qubit or single-spin characteristics. For example, states from the FR ensemble neither display a flat nor a non-flat spectrum, as a single non-zero RDM eigenvalue characterizes them. Nonetheless, their magic originates solely from single-qubit rotations. Quench dynamics induce an entanglement increase, and consequently, this makes the anti-flatness a valuable tool for probing the magic content of the evolving states. 

Examining the anti-flatness in Fig.~\ref{fig:TFIM} (lower panels) reveals a clear separation between integrable and non-integrable dynamics. Additionally, the logarithmic anti-flatness $ \mathrm{F} (\rho_{R})$ similarly exhibits a distinct difference between the two dynamics. Unlike entanglement and magic, where the universal value is approached from below, the anti-flatness approaches the universal value from above (not shown in the given plots). The inclusion of the absolute value in the definition of Eq.~\eqref{relative_difference} thus ensures a positive quantity, enabling an unambiguous differentiation between the dynamics.

\begin{figure}[t!]
 		\centering \includegraphics[width=\columnwidth]{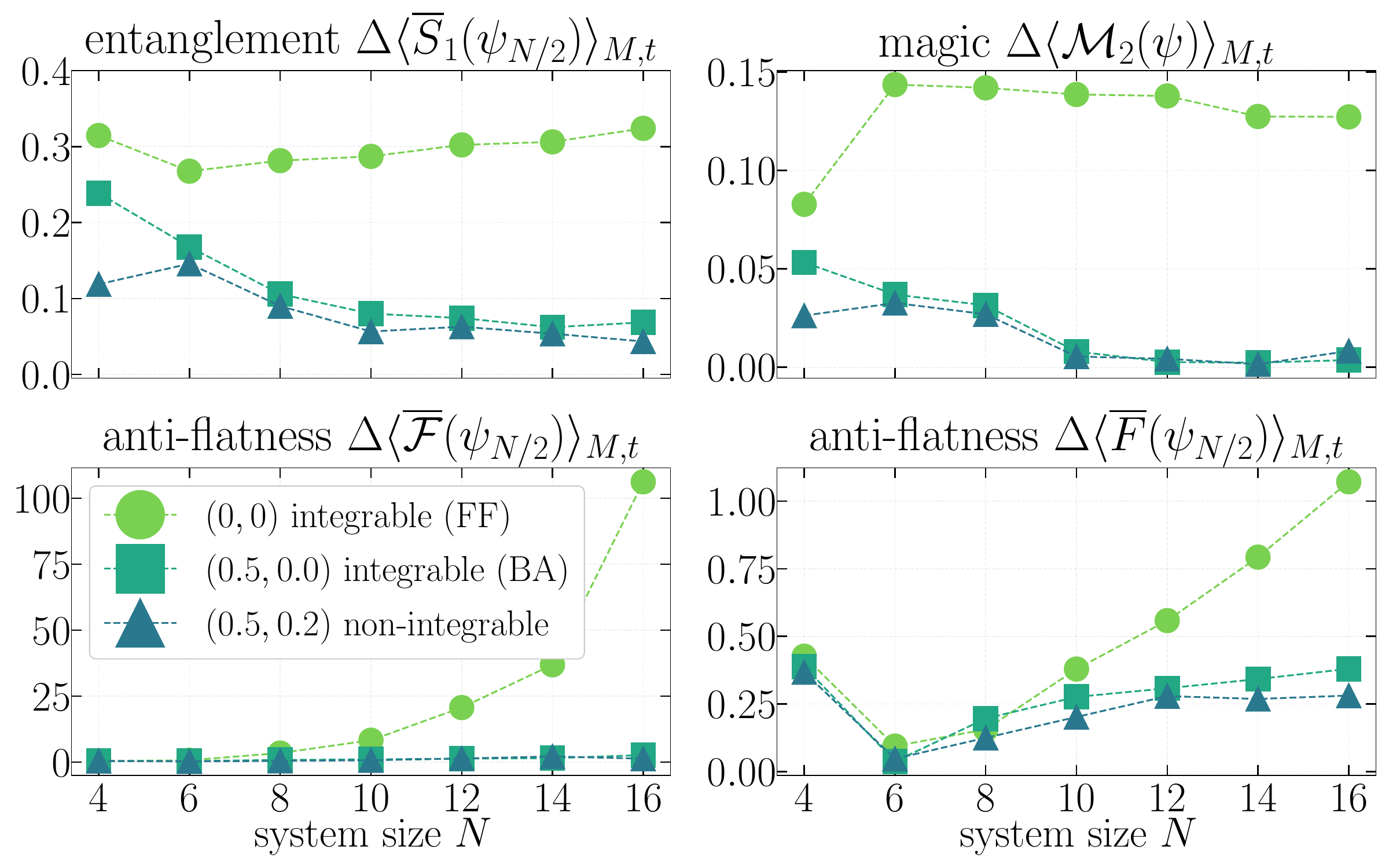}
 		\caption{\justifying Ensemble averaged relative difference defined for different quantities in the case of the long-time quench dynamics generated by the XXZ+NNN Hamiltonian where in brackets the parameters $(\Delta, \alpha)$ are denoted. Parameters choices: $M = 50$, $\Delta t = 2$,   $t^{\rm final} = 10^{4}$ (long-time limit), and FR as the initial state ensemble.
   }
 \label{fig:XXZ}
 \end{figure}

Beyond free-fermion (FF) theories, as a final result of our work, we extend this investigation to models that exhibit integrability with non-trivial interactions, specifically Bethe Ansatz (BA) integrable systems. Quench dynamics in such settings have been extensively studied; see, for instance, Refs.~\cite{Alba_Calabrese_2018, Kormos_Collura_Takacs_Calabrese_2017, Calabrese_2020, DAlessio_Kafri_Polkovnikov_Rigol_2016} and references therein. A prototypical example of a BA integrable model is the XXZ chain, described by the Hamiltonian in Eq.~\eqref{XXZ_Hamiltonian} when the next-to-nearest-neighbor (NNN) coupling vanishes, i.e., $\alpha = 0$. This model is FF integrable for $(\Delta, \alpha) = (0, 0)$ and BA integrable for any $(\Delta, 0)$. Our findings indicate that the long-time behavior of global quantum quench dynamics for this Hamiltonian (shown in Fig.~\ref{fig:XXZ}) reveals trends similar to those observed in the Ising chain. Specifically, Hamiltonian dynamics governed by free fermions do not exhibit the universality and ergodicity characteristic of quantum chaotic systems, as captured by Haar random states. Breaking FF integrability  enables the time-evolved states to achieve full Hilbert space ergodicity.

Interestingly, we find that the dynamics driven by Bethe Ansatz integrable Hamiltonians can resemble those of non-integrable systems, ultimately approaching the expectations of Haar random states. As previously mentioned in the context of quadratic Hamiltonians and integrable dynamics, the long-time states resulting from global quench dynamics are analogous to those found in the middle of the spectrum of the respective Hamiltonian. Ref.~\cite{LeBlond_Mallayya_Vidmar_Rigol_2019} examined the bipartite entanglement entropy of such typical states in the BA integrable XXZ model, revealing a notable similarity to the free-fermion response. Specifically, the average entanglement entropy of these states depends on the system-to-subsystem ratio $f$. However, this observation pertains to typical states in the zero magnetization sector and does not account for states with varying total magnetization. In contrast, the time dynamics we study involve excitations and high-energy states spanning all available magnetization sectors. Consequently, these dynamics are not constrained to a distinct integrable pattern, reflecting a more complex behavior.

In Appendix~\ref{app:OTOC}, we present an additional set of results on a widely used probe of information scrambling and quantum chaos known as the the out-of-time-ordered correlators (OTOCs)~\cite{Roberts_Yoshida_2017, Xu_Swingle_2024}. We emphasize that all variants of OTOCs quantify operator growth rather than state evolution, making a direct comparison with our earlier results on quench dynamics nontrivial. Nevertheless, our study of the OTOCs does provide additional support for the claims made throughout the main text.

\section{Conclusions and outlook}
In this paper, we studied the dynamical behavior of non-stahilizerness resource measured by SRE after a quantum quench of a spin chain. Since SRE is involved in the onset of quantum chaos in its interplay with entanglement production, one would think that the long-time behavior of SRE depends on whether time evolution is generated by a chaotic Hamiltonian or not. We have shown that, in free-fermion theories,  SRE shows a gap with respect to the Haar-value, therefore signaling the lack of quantum complex behavior. On the other hand, non integrable models show a perfect adherence to the Haar value. To understand the role of both entanglement and stabilizer entropy production we use random initial states that possess either resource, or none. Interplay and correlations between entanglement and SRE are of fundamental importance of the understanding of quantum many-body systems. One way to characterize their joint behavior is through measures of anti-flatness which show how SRE is spread across the system. In this work, we show that also anti-flatness measures are capable of distinguishing free-fermion theories non-integrable systems.

Interestingly, the Bethe-ansatz integrable model does not show a gap in their long-time behavior. This suggests that, while the SRE-gap tells apart free-fermion theories, a complete characterization of quantum chaos requires more refined probes, for instance, looking at temporal fluctuations of SRE and higher moments of anti-flatness~\cite{True_Hamma_2022, Leone_Oliviero_Zhou_Hamma_2021}. 

In perspective, we are interested in propagation aspects of SRE, showing how quantum quenches by local perturbations carry SRE about the system and whether its spreading is ballistic or diffusive. Of course, an interesting question is the role played by either criticality or gaplessness in the time evolution. 
The interplay between SRE and entanglement could be investigated through the lens of non-local magic~\cite{Cao_Cheng_Hamma_Leone_Munizzi_Oliviero_2024}. Finally, we are interested in applying these methods in other quantum many-body systems of interest, e.g., disordered systems with quantum many-body localization.

During the completion of this work, we became aware of a complementary study in Ref.~\cite{Tirrito_Turkeshi_Sierant_2024}. Additionally, recent analytical work in Ref.~\cite{Hou_Cao_Yang_2025} has provided further insight into the typical behavior of entanglement and magic and their interplay which has been the central subject of our work. 

The numerical codes, data, and plotting scripts
employed in this work are available at Zenodo~\cite{odavic_2024_14440651}

\section{Acknowledgments} 
We thank Guglielmo Lami, Lorenzo Campos Venuti, Gianluca Esposito, Daniele Iannotti and Immacolata De Simone for useful discussions. This work was also supported by the PNRR MUR Project No. PE0000023-NQSTI. A.H. and M.V. acknowledge financial support from PNRR MUR Project No. CN 00000013-ICSC. J. O. and M. V. acknowledge computational resources from MUR, PON “Ricerca e Innovazione 2014-2020”, under Grant No. PIR01-00011 (I.Bi.S.Co.). JO acknowledges ISCRA for awarding this project access to the LEONARDO super-computer, owned by the EuroHPC Joint Undertaking, hosted by CINECA (Italy) under the project ID: PQC - HP10CQQ3SR.

\bibliography{refs}

\appendix
\section{Out-of-time-ordered correlators (OTOCs)}\label{app:OTOC}

\begin{figure*}[t!]
  \centering
  \begin{subfigure}[t]{0.49\textwidth}
    \centering
    \includegraphics[width=1.025\linewidth]{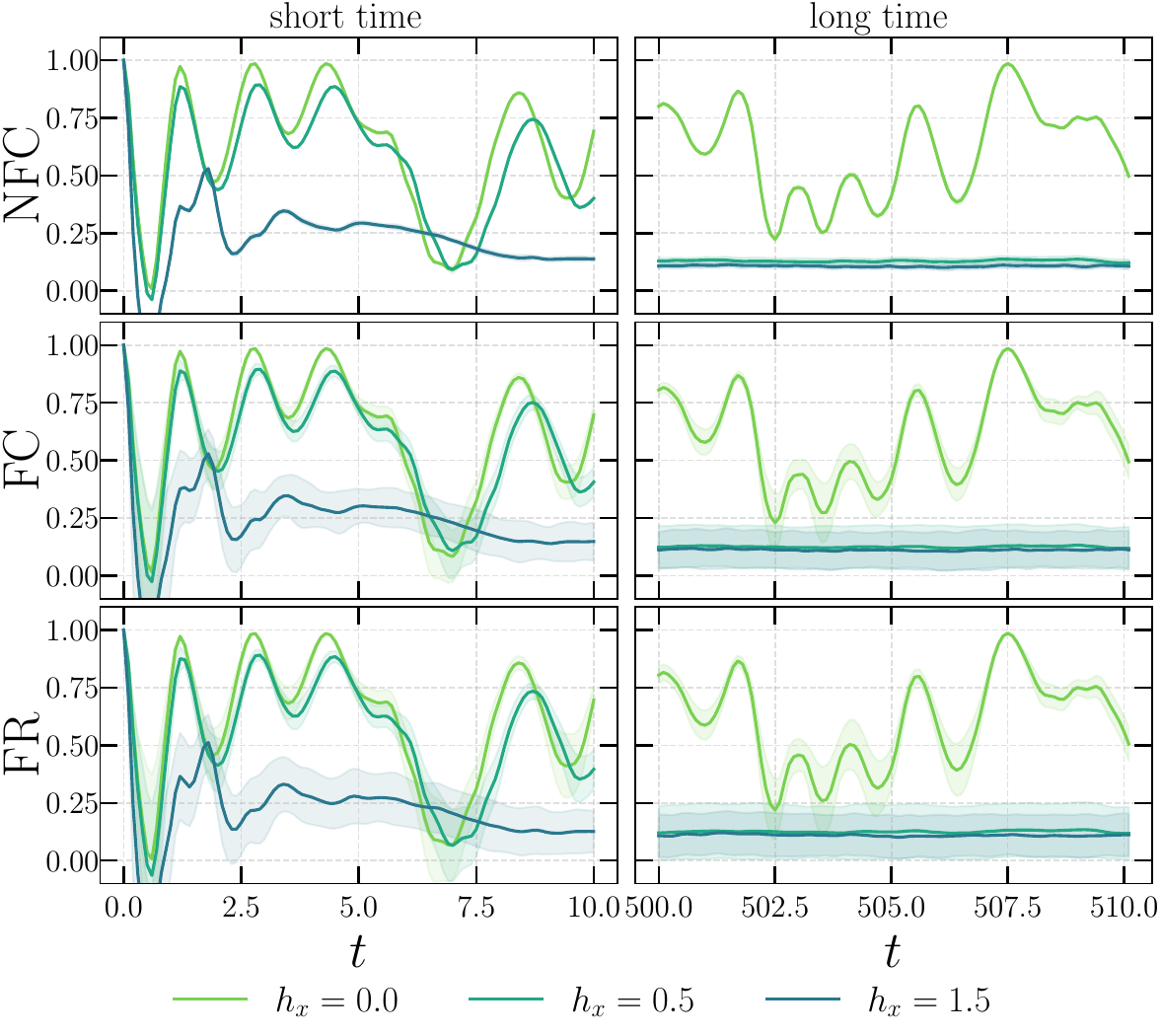}
    \caption{ TFIM+L }
    \label{fig:TFIM_OTOC}
  \end{subfigure}
    \hfill
  \begin{subfigure}[t]{0.49\textwidth}
    \centering
    \includegraphics[width=1.025\linewidth]{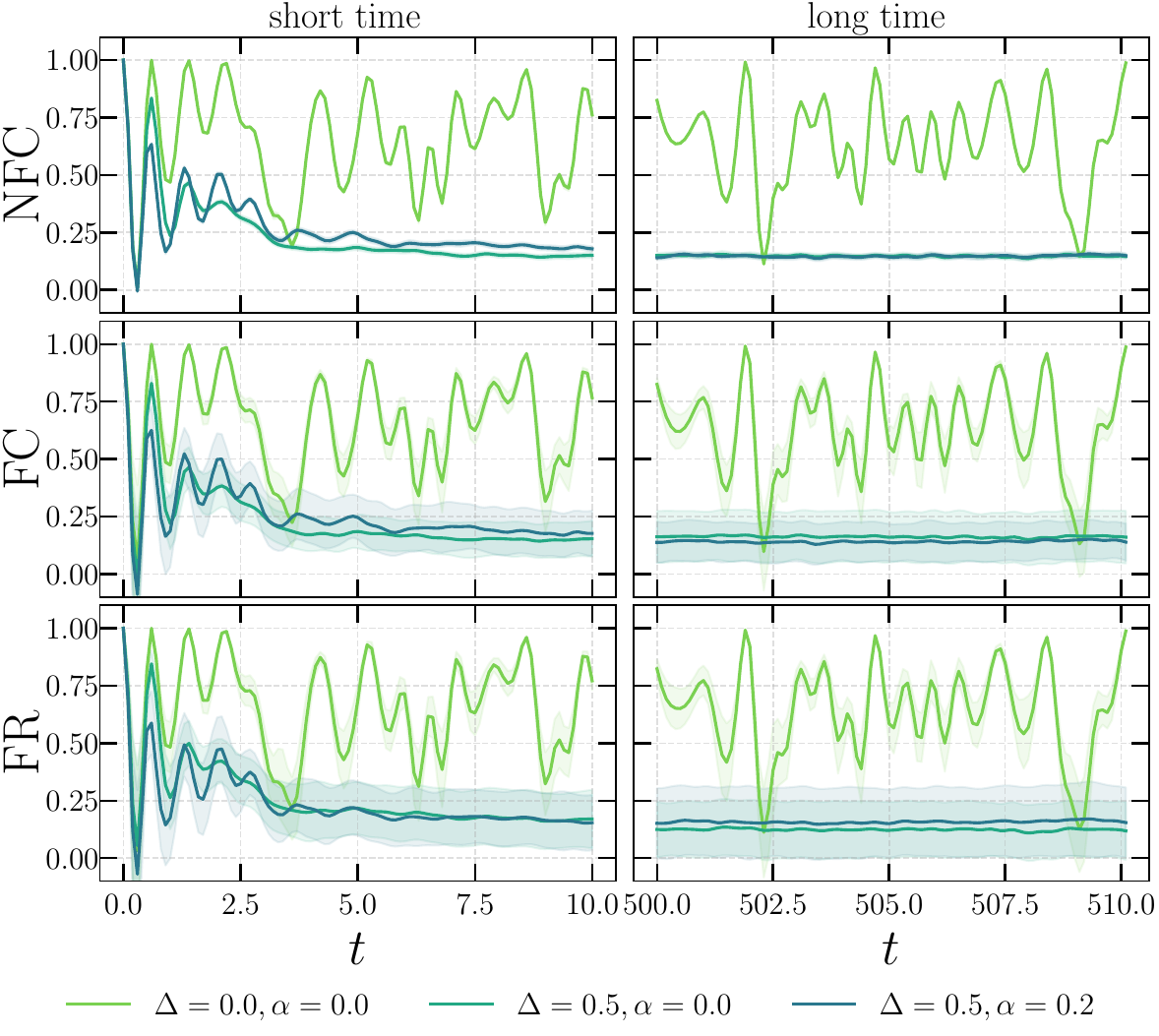}
    \caption{XXZ+NNN}
    \label{fig:zoom}
  \end{subfigure}

  \caption{\justifying Real part of the sample averaged OTOC defined in Eq.~\eqref{eq:OTOC}. The shaded region denotes the standard deviation around this average. System size considered $N = 12$, and for each ensemble we take $M = 100$ realizations. }
  \label{fig:otoc_combined}
\end{figure*}

\noindent We utilize the following definition of the 4-point OTOC, i.e. 
\begin{equation}
    \mathcal{F}(t) = \langle W(t)^\dag V^\dag (0) W(t) V (0) \rangle_\psi,
    \label{eq:OTOC_def}
\end{equation}
where the operators $V(0)$ and $W(0)$ are local, $W(t) = e^{iHt} W(0) e^{-iHt}$, and $\langle O \rangle_\psi = \mathrm{Tr}[ O \psi] $, which, under the assumption that $O$ is Hermitian, can be interpreted as the expectation value of the operator $O$ in the state $\psi$. Much of the existing literature focuses on OTOCs evaluated on the thermal states $\langle \bullet \rangle = {\rm Tr} [e^{- \beta H} \bullet]/ {\rm Tr} [e^{- \beta H }]$. Moreover they are evaluated in infinite temperature limit ($\beta \rightarrow 0$) removing the state dependence from the definition yielding
\begin{align}
   \mathcal{F}_{\beta \rightarrow 0} (t) :=  \frac{1}{d} {\rm Tr} [ W(t)^\dag V^\dag(0) W(t) V(0)].
\end{align}
Given that the main objective of our work is highlighting the state dependence on the time evolution, we keep the state dependence in the definition for the OTOC and draw the states $\psi$ from $\rm NFC$, $\rm FC$ or $\rm FR$ ensembles defined in Sec.~\ref{STS_InitialStates}. These ensemble states $\psi$ are pure, in which case we can simply express the OTOC as
\begin{align}
    \mathcal{F}(t) &= \bra{\psi} W(t)^\dag V^\dag(0) W(t) V(0) \ket{\psi} \notag \\
    &= \braket{\psi_{W(t) V(0)}}{\psi_{V(0) W(t)}}, \label{eq:OTOC}
\end{align}
where
\begin{align}
\ket{\psi_{V(0) W(t)}} &= V(0) W(t)\ket{\psi}, \\
\ket{\psi_{W(t) V(0)}} &= W(t) V(0) \ket{\psi}.
\end{align}
In this form, it becomes evident that the OTOC captures the difference between states obtained by applying operators in different orders. Specifically, it quantifies the extent to which the operators $V(0)$ and $W(t)$ fail to commute when acting on the state $\ket{\psi}$. A straightforward property of the OTOC is that $\mathcal{F}(t) = 1$ whenever $[W(t), V(0)] = 0$. When $V(0)$ and $W(0)$ are supported on spatially disjoint regions, $\mathcal{F}(t)$ remains equal to one until the time evolution causes the support of $W(t)$ to overlap with that of $V(0)$.

As in the main text, we consider spin chains of length $N$, with Hamiltonians given by either TFIM+L or XXZ+NNN, as introduced in Sec.~\ref{sec:description}. We assume periodic boundary conditions (PBC) for the respective Hamiltonians and study the time evolution of the OTOC in both integrable and non-integrable parameter regimes. We take $V(0)$ and $W(0)$ to be Pauli-$z$ operators $\sigma_i^z$ acting on the same $i$-th spin of the chain. When choosing spatially separated local operators we note that the distance between the operators does not affect our conclusions about the infinite-time behavior of the OTOCs. Separating the operators in this way simply extends the time interval during which the OTOC remains equal to one. Moreover, by setting $V = W$, we directly probe how sensitively the initial operator fails to commute with itself under time evolution. Having the setting clearly defined, we ask and answer the following questions
\begin{enumerate}
    \item Are OTOCs sensitive to the choice of the initial family of states?
    \item Can OTOCs distinguish integrable from chaotic Hamiltonians?
\end{enumerate}

To address the first question, Fig.~\ref{fig:otoc_combined} presents the sample-averaged real part of the OTOC, calculated over multiple realizations of the initial state. The imaginary part, which fluctuates around zero on average, is smaller in magnitude and contains less relevant information compared to the real part. The shaded regions represent the standard deviation around the sample mean across $M$ realizations. We find that, at any fixed time $t$, the average OTOC is largely insensitive to the choice of the initial state ensemble. However, a significant difference emerges in the ensemble variance. In both integrable and non-integrable regimes, the FC and FR initial state ensembles exhibit substantially higher variances compared to the NFC ensemble. This trend holds for both of the models considered, including the case where the quench Hamiltonian is Bethe Ansatz integrable. These findings are consistent with the variances observed for the entanglement entropy in Fig.~\ref{fig:short1} and the stabilizer entropy in Fig.~\ref{fig:short2}.

Turning to the second question, which concerns the sensitivity of the OTOC to integrable versus non-integrable dynamics, we again refer to Fig.~\ref{fig:otoc_combined}. In the long-time limit, where the system approaches a steady state, the distinction between integrable and non-integrable dynamics becomes pronounced. Interestingly, the behavior of the OTOC following a Bethe Ansatz integrable quench appears similar to that of a non-integrable (chaotic) system, consistent with the trends observed using other probes discussed in the main text. It is well established that in chaotic systems, OTOCs saturate at late times~\cite{Fortes_2019,Heyl_2018}, whereas integrable systems often show persistent oscillations. Our results are fully aligned with these expectations for both the TFIM+L and XXZ+NNN models. As we tune the parameters away from integrable points and into the chaotic regime, the onset of saturation occurs progressively earlier, especially when the integrability-breaking parameter is large. Moreover, although not shown in the figure, we observe that the late-time saturation values for both the non-integrable and Bethe Ansatz integrable quenches are remarkably close to the values expected from a Haar-random ensemble of initial states.

Drawing inspiration from Ref.~\cite{Roberts_Yoshida_2017}, particularly Sec.~5 on the behavior of the OTOC, we belive that the operator dynamics observed in our work may be further understood by analyzing the frame potentials thereby investigating the degree to which our initial state ensembles approximate $t$-designs.

\end{document}